\documentclass[conference]{IEEEtran}
\IEEEoverridecommandlockouts
\usepackage{cite}
\usepackage{amsmath,amssymb,amsfonts}
\usepackage{algorithmic}
\usepackage{graphicx}
\usepackage{textcomp}
\def\BibTeX{{\rm B\kern-.05em{\sc i\kern-.025em b}\kern-.08em
    T\kern-.1667em\lower.7ex\hbox{E}\kern-.125emX}}

\usepackage{url}
\usepackage{verbatim}
\usepackage{color}
\usepackage{enumitem}
\usepackage{boxedminipage}
\usepackage{microtype}
\usepackage[caption=false,font=footnotesize]{subfig}
\usepackage{eso-pic}

\setlist{leftmargin=*, noitemsep}

\mathchardef\mhyphen="2D  

\newcommand{\Bytes}{\mathsf{Bytes}}
\newcommand{\Struct}{\mathsf{Struct}}
\newcommand{\Array}{\mathsf{Array}}
\newcommand{\List}{\mathsf{List}}

\newcommand{\Num}{\mathsf{Num}}
\newcommand{\Type}{\mathsf{Type}}
\newcommand{\ChildPtr}{\mathsf{ChildPtr}}
\newcommand{\NextPtr}{\mathsf{NextPtr}}
\newcommand{\NULL}{\mathsf{NULL}}
\newcommand{\VisitNode}{$N_v$}
\newcommand{\VisitField}{$F_v$}
\newcommand{\TopContext}{$C_{top}$}
\newcommand{\ListLevel}{\mathsf{ListLevel}}

\begin{document}
    
\bstctlcite{ref:BSTcontrol}

\title{HGum: Messaging Framework for Hardware Accelerators}

\author{\IEEEauthorblockN{Sizhuo Zhang}
    \IEEEauthorblockA{MIT CSAIL\\
        szzhang@csail.mit.edu}
    \and
    \IEEEauthorblockN{Hari Angepat}
    \IEEEauthorblockA{Microsoft\\
        Hari.Angepat@microsoft.com}
    \and
    \IEEEauthorblockN{Derek Chiou}
    \IEEEauthorblockA{Microsoft\\
        dechiou@microsoft.com‎}}

\maketitle

\begin{abstract}
	
Software messaging frameworks help avoid errors and reduce engineering
efforts in building distributed systems by (1) providing an interface
definition language (IDL) to specify precisely the structure of the
message (i.e., the message {\em schema}), and (2) automatically generating
the serialization and deserialization functions that transform user
data structures into binary data for sending across the network and
vice versa.  Similarly, a hardware-accelerated system, which consists of
host software and multiple FPGAs, could also benefit from a messaging
framework to handle messages both between software and FPGA and also
between different FPGAs.  The key challenge for a hardware messaging
framework is that it must be able to support large messages with
complex schema while meeting critical constraints such as clock
frequency, area, and throughput.  

In this paper, we present HGum, a messaging framework for hardware
accelerators that meets all the above requirements.  HGum is able to
generate high-performance and low-cost hardware logic by employing a
novel design that algorithmically parses the message schema to perform
serialization and deserialization.  Our evaluation of HGum shows that
it not only significantly reduces engineering efforts but also
generates hardware with comparable quality to manual implementation.

\end{abstract}


\section{Introduction}

Messaging frameworks are very useful in building distributed software systems.
In such systems, processes on different servers communicate by sending messages over the network.
Although libraries exist to handle networking (e.g., Ethernet), there is still a need to encode a message into the binary format (e.g., a blob of bytes) that is sent to the network, and decode the received binary data (from network) to re-construct the message.
Since different processes may be developed by different teams, the \emph{schema} (i.e., the structure) of the message and the encoded binary format must be documented so that all teams can generate a consistent encoding and decoding of the message.
Such documents are generally written in English rather than as a formal specification.

Hand implementations of encode/decode functions are tedious and error-prone.
Whenever the message schema is updated, the functions must be re-implemented.
Besides, documents of the schema can be interpreted differently by two teams which may use different programming languages.
Such inconsistency may cause the data structure (in each language) that represents the message to not match the defined schema.

\begin{figure}[!htb]
    \vspace{-10pt}
	\centering
	\includegraphics[width=0.7\columnwidth]{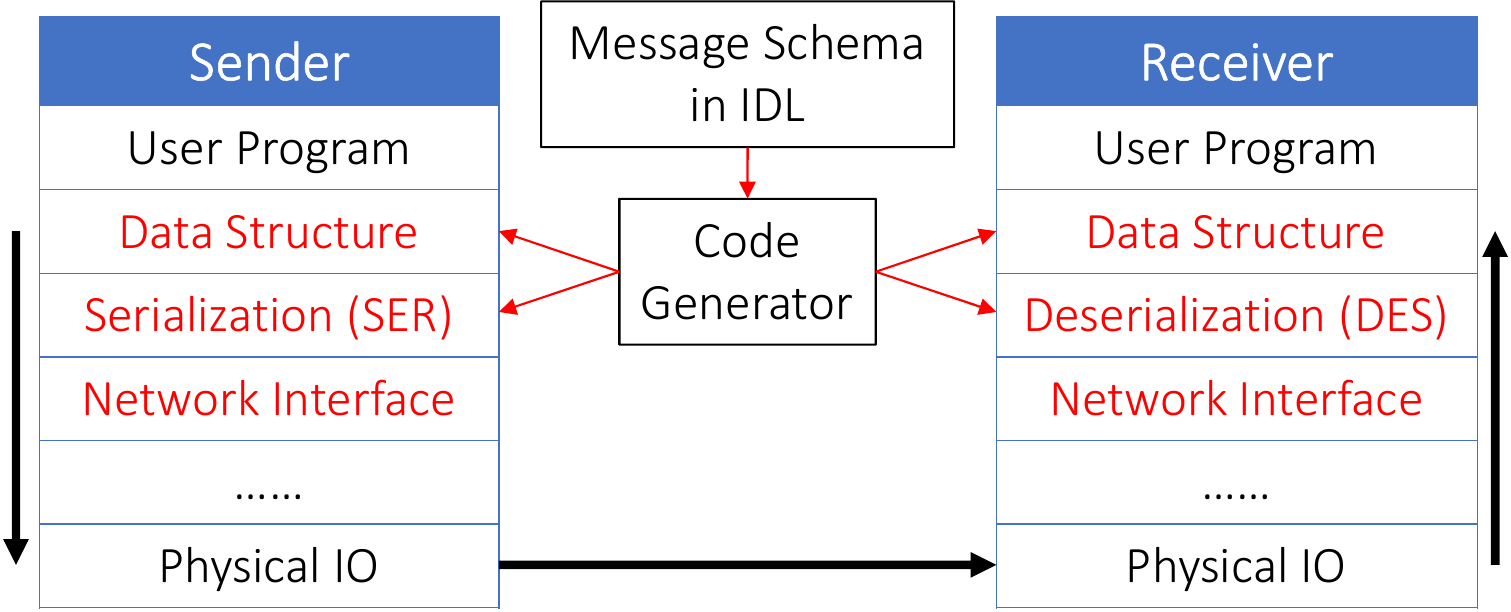}
    \vspace{-5pt}
	\caption{Messaging framework architecture} \label{fig: msg framework arch}
    \vspace{-10pt}
\end{figure}

Various messaging frameworks~\cite{protobuf,bond,thrift,capnproto,flatbuffers}
have been developed to automatically generate codes to encode/decode messages.
These frameworks share the architecture shown in Figure~\ref{fig: msg framework arch}.
The sender composes a data structure that represents the message, and then calls the \emph{serialization} (SER) function to encode the data structure into binary data, which is sent to the receiver via network.
The receiver uses the \emph{deserialization} (DES) function to decode the binary data and re-construct the data structure.

The network interface in the messaging framework abstracts away the details of the network and
permits the use of different network protocols.  More importantly,
the framework automatically generates the data structures and
SER/DES functions based on the message
schema, which is unambiguously specified by the user in an
\emph{interface definition language} (IDL).  These features of the
messaging framework address all the previously mentioned problems.

Recently, FPGAs have been used in accelerating many services.
For example, Microsoft has used an 8-FPGA pipeline to
accelerate Bing search~\cite{putnam2014reconfigurable}.
In this system, messages are exchanged not only between a software host
and an FPGA, but also between FPGAs.  The messages can be large and complex data structures.
For instance, a request message from the software host to the FPGA
pipeline can be 64-KB large, and consist of multiple levels
of nested arrays.  To reduce the complexity and errors introduced by hand coding, we propose \textbf{HGum}, a messaging
framework for hardware accelerators.

\begin{figure}[!htb]
	\centering
	\includegraphics[width=0.7\columnwidth]{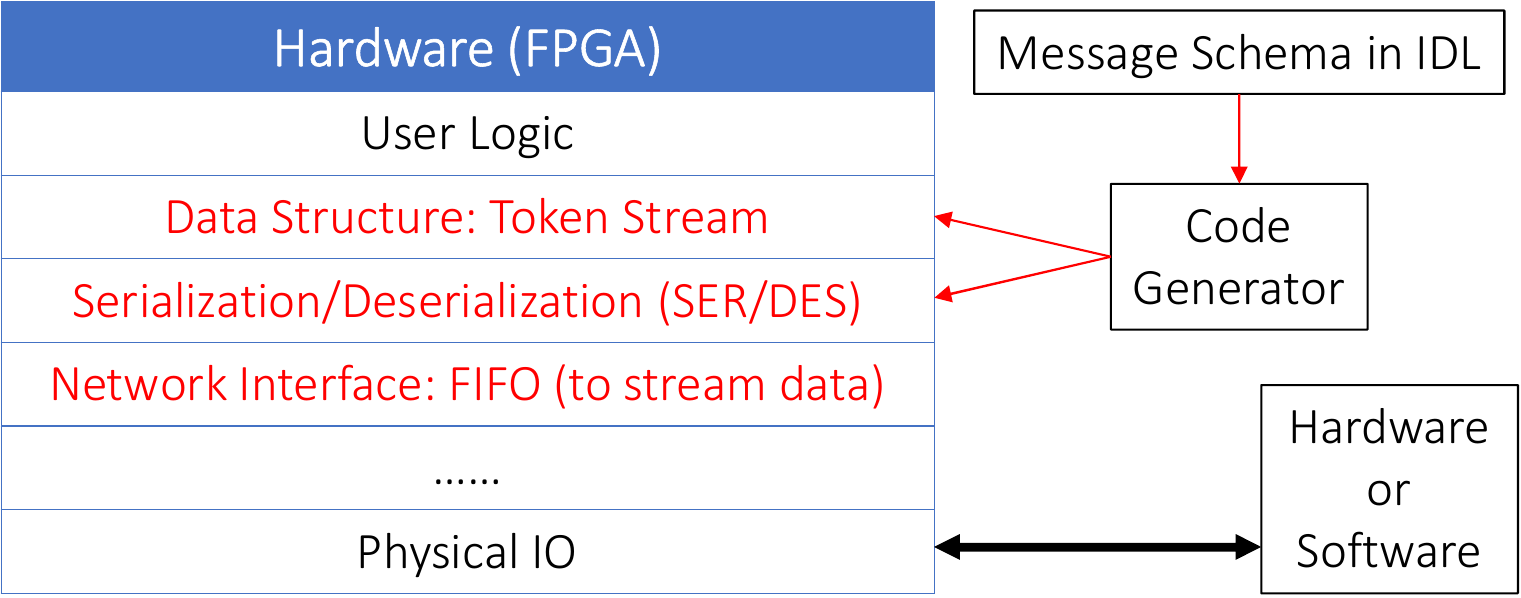}
    \vspace{-5pt}
	\caption{HGum framework architecture} \label{fig: hgum stack}
    \vspace{-15pt}
\end{figure}

Figure~\ref{fig: hgum stack} shows the architecture of HGum,
which is similar to that of a software messaging framework.
The difference is that HGum allows hardware to exchange
messages with both software and other hardware.  HGum faces
the following challenges not present for software messaging frameworks:
\begin{enumerate}
	\item Limited hardware resources: Since the message can be
	very large, it cannot be buffered on chip entirely, nor can the
	port be the width of the entire message.  Therefore the
	SER/DES logic must process data in a \emph{streaming} fashion.
	
	\item Performance and area constraints: 
    The SER/DES logic should not be the bottleneck of the accelerator design.
    Thus the generated SER/DES logic must have high throughput, run at high
	frequency, and occupy as little area as possible.
\end{enumerate}

To address the first challenge, the network interface in HGum is
defined as a pair of FIFO interfaces to stream in and out fixed-width data,
and the ``data structure" exchanged between user logic and SER/DES
logic is defined as a stream of \emph{tokens} for incremental
composition and decomposition of the message. We refer to the fixed-width data
at the network interface as \emph{phit}, and its
width is often set to match the bandwidth of physical links.  Each
token in the token stream can be viewed as a lowest-level field of the
message.  It contains structural information about its location
inside the message schema.  Such information is generated by the
DES logic to enable user logic to easily make use of the
tokens.  The streaming network interface and the token-stream
``data structure" distinguish HGum from software messaging frameworks,
where the network interface is typically a randomly accessible
buffer and the data structure contains the whole message.

The second challenge is addressed by a novel design of the SER/DES
logic, which parses the message schema in an algorithmic way.
This approach makes the performance and area of the SER/DES logic almost insensitive to the
complexity of the message schema.
Therefore, HGum provides strong guarantees on performance and
area.  

To the best of our knowledge, HGum is the first messaging framework
for hardware to process large and complex messages in a streaming fashion.

\noindent\textbf{Paper organization:}
Section \ref{sec: relate work} discusses other messaging
frameworks for hardware.
Section \ref{sec: spec} specifies the interfaces and the workflow of HGum.
Section \ref{sec: impl} elaborates the implementation of HGum.
Section \ref{sec: eval} evaluates HGum for both basic message schema and complex schema used in real accelerators.
Section \ref{sec: conclude} offers the conclusion.

\section{Related Work} \label{sec: relate work}

SER/DES functions in software messaging frameworks
\cite{protobuf,bond,thrift,capnproto,flatbuffers} are store-and-forward.
That is, deserialization will not start before all the binary data is received
from the network, and the serialized data will not be passed to the
network until serialization is fully completed.  In contrast, HGum
processes messages in the streaming fashion, i.e., the
SER/DES logic can function in parallel with the network transport layer.

There is a plethora of prior work on hardware messaging frameworks.
CoRAM~\cite{chung2011coram} bridges FPGA and external memory by
allocating cache-like storage on FPGA.  Users can program the control logic of
on-FPGA caches in a C-like language.
Unlike HGum, CoRAM is unaware of the data schema transferred between
memory and on-FPGA caches, so the user still needs to handle the binary data in on-FPGA caches.

LEAP~\cite{fleming2014leap} supports FIFO interfaces for messaging,
but the FIFO width is equal to the message size.
This incurs huge area costs when transferring large messages.
In contrast, HGum can transfer messages of arbitrarily large sizes.

A large number of the remaining hardware messaging frameworks only target
systems with a host processor and a single FPGA.
Many of these frameworks~\cite{king2013generating,king2015software,willenberg2013remote} make the communication between the host processor and FPGA easier.
Other instances~\cite{peck2006hthreads,huang2008liquid,filgueras2014ompss,xilinxOpenCL,alteraOpenCL} provide higher-level FPGA programming
languages.  All such frameworks only support communication between
hardware and software, while HGum also supports messaging from
hardware to hardware.

\section{Specification of HGum} \label{sec: spec}

\subsection{Informal Example of SER/DES}

Before giving the formal specifications, we use two examples to illustrate the SER/DES functionality generated by HGum.

\begin{figure}[!htb]
	\centering
    \vspace{-5pt}
	\includegraphics[width=0.9\columnwidth]{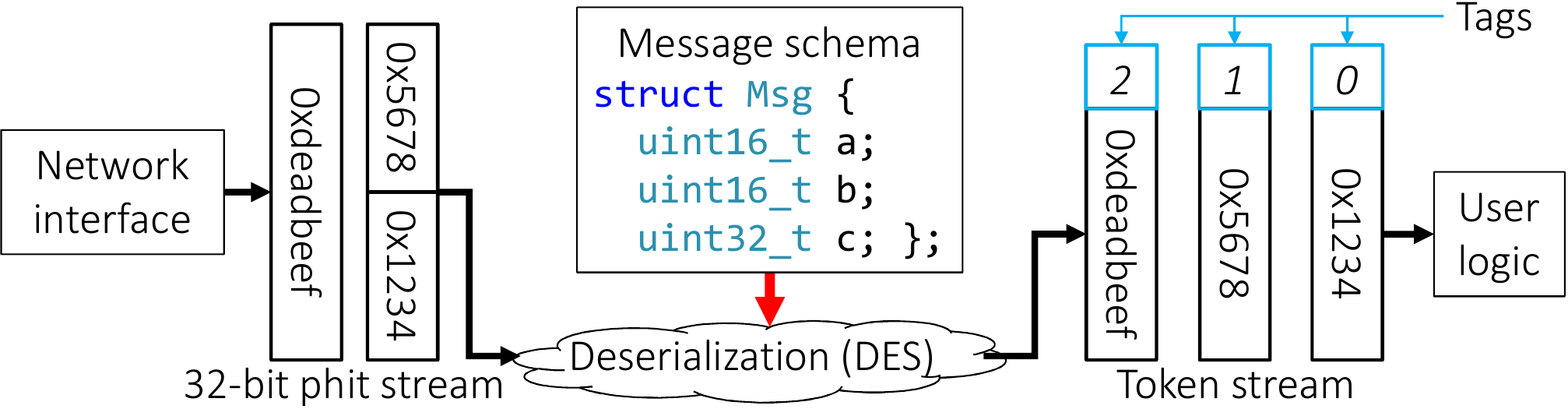}
    \vspace{-5pt}
	\caption{Simple example for deserialization (DES)} \label{fig: des simple ex}
    \vspace{-5pt}
\end{figure}

\noindent\textbf{Deserialization (DES):}
Figure~\ref{fig: des simple ex} shows an example of deserializing a
phit stream from the network to a stream of tokens that will be
sent to the user logic. The message schema is informally
represented by a C++ structure. In this example, a phit is 32-bit
wide. The first phit contains the first two message fields, $a=$0x1234 and $b=$0x5678, and the second phit contains the
last field $c=$0xdeadbeef. The DES logic outputs
each field as a token in the same order that the fields appear in the
message schema. Each token has a tag, which is informally
shown as 0, 1 and 2 respectively in Figure~\ref{fig: des simple ex}.
The tag indicates the field that the token corresponds to, and helps the user logic consume the token.
Section~\ref{sec: des token spec} will show how a user can specify the tag associated with each token in the HGum IDL.

\begin{figure}[!htb]
	\centering
    \vspace{-5pt}
	\includegraphics[width=0.9\columnwidth]{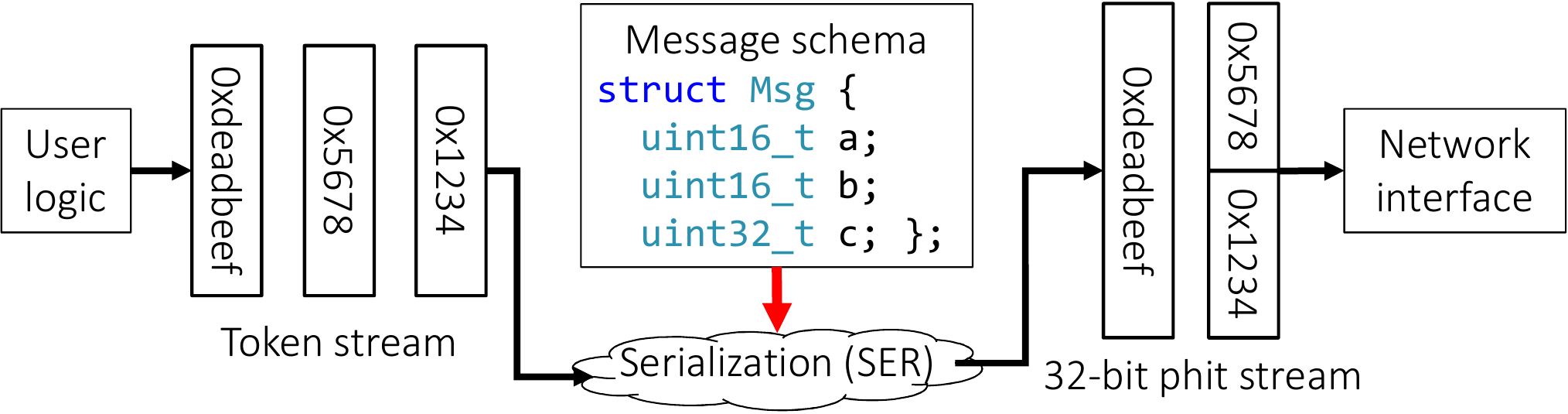}
    \vspace{-5pt}
	\caption{Simple example for serialization (SER)} \label{fig: ser simple ex}
    \vspace{-10pt}
\end{figure}

\noindent\textbf{Serialization (SER):}
Figure~\ref{fig: ser simple ex} shows an example of serializing a token stream from the user logic to a stream of phits.
This is almost the reverse of the DES example in Figure~\ref{fig: des simple ex} except that, in this case, tokens do not have tags.
The SER logic does not need tags to determine the field of each token in the schema, because the tokens are emitted by the user logic in the same order as the fields appear in the message schema.

\subsection{Formal Specification of Schema IDL}

HGum uses JSON as the IDL to specify message schema.
Figure~\ref{fig: idl grammar} shows the grammar of the HGum IDL in JSON.

\begin{figure}[!htb]
	\centering
	\begin{boxedminipage}{\columnwidth}
		\small \vspace{-10pt}
		\begin{eqnarray}
		schema & ::= & \{ \ structName : structDef, \nonumber \\
		&     & \  \ structName : structDef, \ldots \ \} \nonumber \\
		structDef & ::= & [\ [fieldName, type], [fieldName, type], \ldots ] \nonumber \\
		type & ::= & [ \Bytes, n ] \ ||\ [ \Struct, structName ] \ || \nonumber \\ 
		&     & [ \Array, type ] \ ||\ [ \List, type ] \nonumber
		\end{eqnarray}
	\end{boxedminipage}
	\caption{Grammar of HGum IDL to define schema} \label{fig: idl grammar}
    \vspace{-15pt}
\end{figure}

In Figure~\ref{fig: idl grammar}, the message schema definition ($schema$) is a collection of structures, i.e., a mapping from structure names ($structName$) to the corresponding definitions ($structDef$).
The $structName$ of the top level structure should match the name of the message.
The structure definition ($structDef$) is an ordered list of fields, and each field is a tuple of field name ($fieldName$) and its type ($type$).
The order of fields in the structure definition determines the order of tokens in the token stream.
HGum supports four types as shown in the definition of $type$ in Figure~\ref{fig: idl grammar}, and their meanings are listed below:
\begin{itemize}
	\item $[ \Bytes, n ]$: $n$-byte data.
	One byte is at default 8-bit wide, but the user can configure its width.
	
	\item $[ \Struct, structName ]$: a structure with name $structName$.
	
	\item $[ \Array, type ]$: an array of elements.
	The element type is $type$.
	The array length is \emph{known} before any element is serialized.
	
	\item $[ \List, type ]$: a list of elements of type $type$.
	The list length remains \emph{unknown} until the last element is serialized.
\end{itemize}
Most software messaging frameworks only have $\Array$ type and do not have $\List$ type.
This is because
software can buffer the whole message before serialization starts, and the size of a linear container is always known.
However, in case of hardware, since we cannot buffer the whole message, we need the $\List$ type when the container size is not known in advance.

The above definitions allow arbitrary nesting of $\Array$, $\List$ and $\Struct$ types.
This is one of the powerful features of HGum.

\begin{figure}[!htb]
    \vspace{-5pt}
	\centering
	\includegraphics[width=\columnwidth]{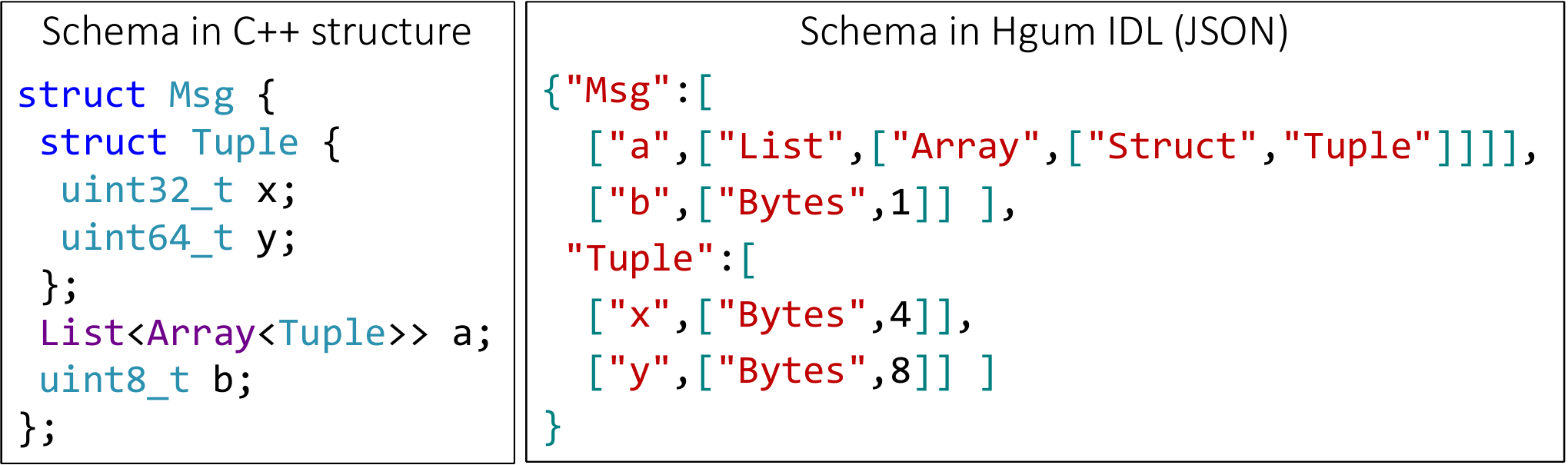}
    \vspace{-15pt}
	\caption{Example of HGum's IDL} \label{fig: idl example}
    \vspace{-10pt}
\end{figure}

As an example, Figure~\ref{fig: idl example} shows the definition of message schema $Msg$ in HGum's IDL. 
(The schema is informally represented by a C++ structure at the left.)
Note that the schema specified in Figure~\ref{fig: idl example} is shared by both the sender and the receiver of the message, so we refer to it as \emph{central schema}.

\subsection{Formal Specification of Tokens}

The token stream is an in-order representation of all the fields in the message, so we define these tokens by recursively translating each field of the top-level structure in the message schema into tokens.
As we have noted above, the tokens outputted from the DES logic take a different format than that of the tokens sent into the SER logic.
This is because we try to encode as much information as possible in the tokens outputted from DES logic to facilitate user logic in consuming the tokens, and we also try to require as less information as possible in the tokens sent into SER logic to reduce the burden on user logic which generates these tokens.
Therefore we specify these two kinds of tokens separately.

\subsubsection{Tokens Outputted From DES Logic} \label{sec: des token spec}

Below we list the tokens (outputted from DES logic) that a field of the message structure corresponds to, depending on the type of the field.
\begin{itemize}
	\item $[ \Bytes, n ]$: A single token containing the $n$-byte data.
	
	\item $[ \Struct, structName ]$: This field can be translated to a series of tokens, which is the concatenation of tokens corresponding to each sub-field of this structure $structName$.
	
	\item $[ \Array, type ]$: This field can be translated to an array-length token followed by the concatenation of tokens corresponding to each element of the array.
	An array-end token can be optionally appended at the end.
	
	\item $[ \List, type ]$:  This field can be translated to a list-begin token followed by the concatenation of tokens corresponding to each element of the list, and finally a list-end token.
\end{itemize}
The tokens for $\Bytes$ and $\Struct$ types are straightforward.
As for the $\Array$ type, the array-length token contains the number of elements in the array, and the optional array-end token can save the effort of user logic to keep track of the end of the array.
We will later introduce how user logic can specify whether an array-end token should be appended.
As for the $\List$ type, since list size is unknown in advance, we use a list-begin token and a list-end token to indicate the begin and end of the list.

We use the message schema in Figure~\ref{fig: idl example} as an example.
When list $a$ has one element and the inner array has two elements, the DES logic will output the following token stream:\vspace{-3pt}
\begin{displaymath}
\begin{array}{c}
a.\mathsf{list\mhyphen{}begin} \rightarrow a[0].\mathsf{array\mhyphen{}length} \rightarrow a[0][0].x \rightarrow a[0][0].y \\ \rightarrow a[0][1].x \rightarrow a[0][1].y \rightarrow a[0].\mathsf{array\mhyphen{}end} \rightarrow a.\mathsf{list\mhyphen{}end} \rightarrow b
\end{array}
\vspace{-3pt}
\end{displaymath}
The first token ($a.\mathsf{list\mhyphen{}begin}$) is the list-begin token for $a$.
Since the element of $a$ is an array, the DES logic then outputs the array-length token for $a[0]$ (i.e., first element of list $a$), followed by the elements of array $a[0]$ ($a[0][0].x \sim a[0][1].y$) and the array-end token ($a[0].\mathsf{array\mhyphen{}end}$).
The list-end token for $a$ ($a.\mathsf{list\mhyphen{}end}$) is outputted afterwards.
Finally, the token ($b$) for field $b$ is outputted.

\begin{figure}[!htb]
    \vspace{-10pt}
	\centering
	\includegraphics[width=0.4\columnwidth]{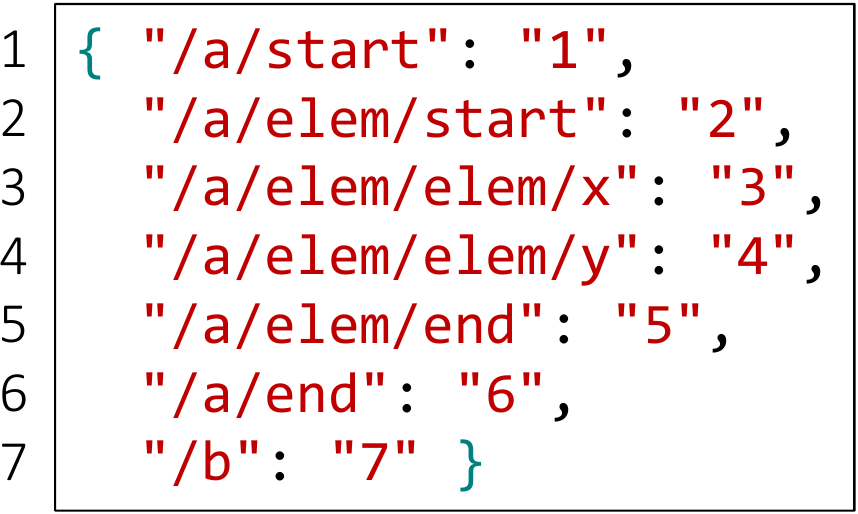}
    \vspace{-5pt}
	\caption{Example of specifying tags for message schema in Figure \ref{fig: idl example}} \label{fig: client schema example}
    \vspace{-15pt}
\end{figure}

As shown in Figure~\ref{fig: des simple ex}, each token from the DES logic has a tag.
The tag can be specified by the user in JSON as follows:\vspace{-3pt}
\begin{displaymath}
\{ path\ of\ the\ token\ in\ the\ message\ struct : tag\ value, \ldots \}
\vspace{-3pt}
\end{displaymath}
Figure~\ref{fig: client schema example} shows the JSON specification of tags for the message schema in Figure~\ref{fig: idl example}.
The first line defines the tag of token $a.\mathsf{list\mhyphen{}begin}$ to be 1, where the keyword ``start'' denotes the array-length or list-begin token.
The third line defines the tags of tokens $a[i][j].x$ (for all $i$ and $j$) to be 3.
The keyword ``elem" refers to the element of an array or list, and ``x'' is the field name.
The fifth line specifies the tags of $a[i].\mathsf{array\mhyphen{}end}$ (for all $i$) to be 5.
The keyword ``end" refers to the array-end or list-end token.
Since the tags of array-end tokens are defined, the DES logic will output these tokens.
To stop emitting these array-end tokens, we can simply remove the fifth line.

It should be noted that the user can specify multiple different tagging schemes for a single message schema, and each tagging scheme is associated with one unique DES module.
Therefore, we refer to the tagging scheme as \emph{client schema}.

\subsubsection{Tokens Sent into SER Logic}

The translation from message structure fields to tokens sent into SER logic is similar to that of DES logic except for the following differences:
\begin{enumerate}
	\item There is no array-end token for the $\Array$ type, because the SER logic can keep track of the end of array.

	\item There is no list-begin token for the $\List$ type.
	Instead, the list-end token will contain the nesting level for lists.
	For example, if a field has type $[\List,[\List,[\Bytes,4]]]$, the list-end token of its inner list should have nesting level 2.

	\item There is no tag for any token, because the SER logic can get enough information from the central schema.
\end{enumerate}

\subsection{Software SER/DES Functions}

The software interfaces of the SER/DES functions in HGum are the same
as those in software messaging frameworks.  The network
interface is a randomly accessible buffer, and the data structure
exchanged between SER/DES functions and user code is the whole message
structure.  The SER/DES functions use the store-and-forward policy to
process messages.

\subsection{Workflow of HGum}

The workflow of HGum consists of the following four steps:
\begin{enumerate}
	\item Define the central schema of the message.
	\item Define one client schema for each hardware DES module.
	\item  HGum automatically generates the SER/DES logic.
	HGum supports passing messages from software to hardware, from hardware to software, and from hardware to hardware.
	The generated software SER/DES functions are in C++, and the hardware SER/DES modules are in SystemVerilog.
	\item Connect the generated SER/DES logic to the network transport layers (e.g., DMA for software-to-hardware messaging).
\end{enumerate}

\section{Implementation of HGum} \label{sec: impl}

\begin{figure}[!htb]
	\centering
	\begin{minipage}{\columnwidth}
		\centering
		\includegraphics[width=0.95\columnwidth]{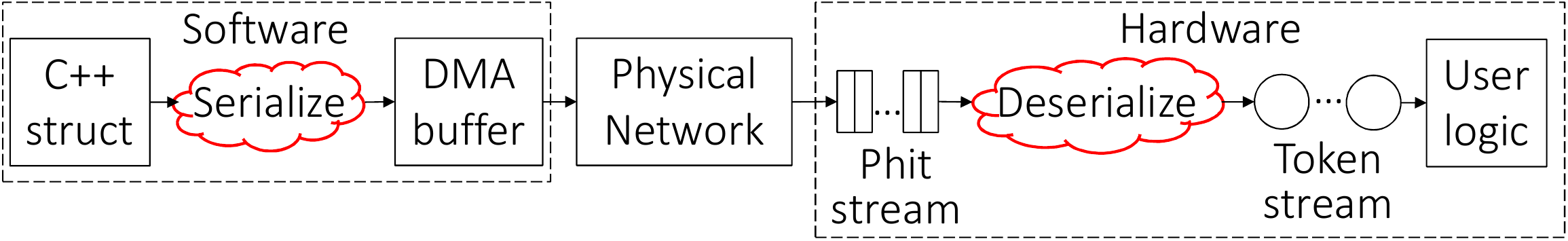}
        \vspace{-5pt}
		\caption{SW-to-HW messaging microarchitecture} \label{fig: sw-hw arch}
	\end{minipage}
	\vspace{8pt}\\
	\begin{minipage}{\columnwidth}
		\centering
		\includegraphics[width=0.95\columnwidth]{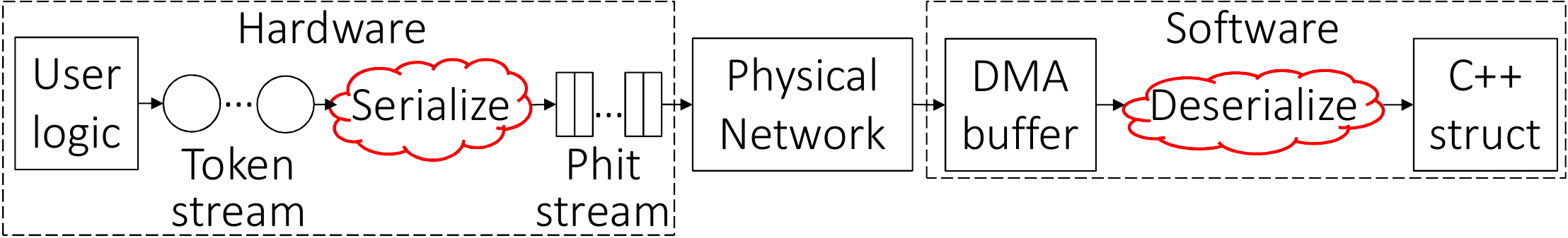}
        \vspace{-5pt}
		\caption{HW-to-SW messaging microarchitecture} \label{fig: hw-sw arch}
	\end{minipage}
	\vspace{8pt}\\
	\begin{minipage}{\columnwidth}
		\centering
		\includegraphics[width=\columnwidth]{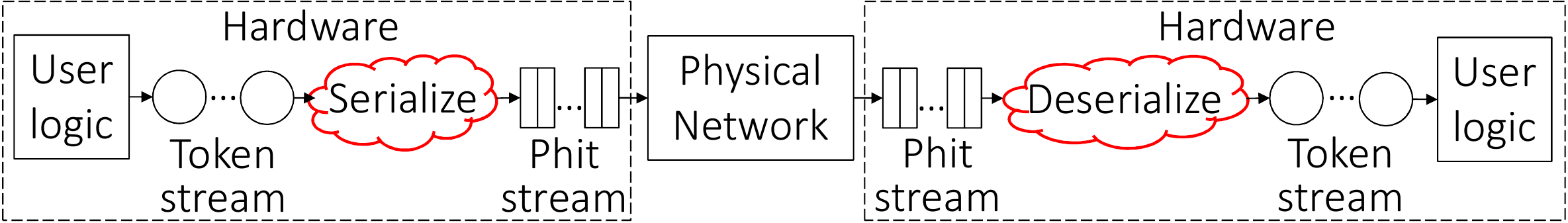}
        \vspace{-15pt}
		\caption{HW-to-HW messaging microarchitecture} \label{fig: hw-hw arch}
	\end{minipage}
    \vspace{-15pt}
\end{figure}

Figures~\ref{fig: sw-hw arch}$\sim$\ref{fig: hw-hw arch} show the
microarchitectures of the HGum framework for all three types of
messaging: software to hardware (SW-to-HW), hardware to software
(HW-to-SW), and hardware to hardware (HW-to-HW).  Since hardware
processes messages in a streaming fashion while software can afford a
store-and-forward policy, the SER/DES logic for the above three types
of messaging are not the same.  We also
leverage the buffering capability to
optimize the designs for SW-to-HW and HW-to-SW messaging.  Next we
detail how to generate the logic for these three
types of messaging respectively.

\subsection{Implementation of SW-to-HW Messaging}

\subsubsection{SER Function in Software}

As shown in Figure~\ref{fig: sw-hw arch}, the software SER function encodes the message into a randomly accessible buffer.
For this, HGum employs a simple binary protocol, which sequentially writes each field of the message into the buffer.
Note that no tag is written to the buffer.
In the protocol, we take the following actions for each field of the message structure based on the type of the field:
\begin{itemize}
	\item $[\Bytes,n]$: directly writes the $n$-byte data to the buffer.
	
	\item $[\Struct,structName]$: serializes each field of the structure into the buffer in the order as the fields appear in the structure.
	
	\item $[\Array,type]$ or $[\List,type]$: first writes the number of elements to the buffer, and then serializes each element into the buffer (from first to last).
	Since software buffers the whole message, $\Array$ and $\List$ are treated in the same way.
\end{itemize}
It is trivial to generate codes for this SER function. 

\subsubsection{DES Logic in Hardware}

According to the protocol of the software SER function, the hardware DES logic only needs to sequentially read each field of the message schema from the phit stream.
The main problem is how to sequentially traverse the message schema.

A naive approach is to use a finite state machine (FSM), in which each state corresponds to a field in the message structure or sub-structure.
However, the number of states in the FSM is roughly equal to the total number of fields in all structures in the message schema, which can be huge if the schema is complex.
Then it would be difficult to provide guarantees on the critical path delay or area. 

\noindent\textbf{Algorithmic traversal of schema:}
Alternatively, HGum employs a novel approach to traverse the schema in an algorithmic way.
The idea is to view the message schema as a tree, which we refer to as \emph{schema tree}. 
Then the traversal of the message schema will become the traversal of the schema tree, which can be done algorithmically using a stack.
The benefit of this approach is that we only need a small FSM, which is \emph{independent} from the message schema, to perform the traversal. 
Therefore,  HGum provides strong guarantees on the critical path delay and area.
In particular, the critical path delay is almost insensitive to the complexity of the message schema.

Before formally defining the schema tree, we first make the following two transformations on the message schema:
\begin{enumerate}
	\item Any array element type or any list element type that is
	not a structure is wrapped into a new $\Struct$ type, making
	the element of every array and list a structure.  
    To simplify description, we assume the structure name of the element type for each array or list is unique.
	
	\item Each structure field which is also of the $\Struct$ type is replaced with its sub-fields.
	That is, every field of a structure can only be of $\Bytes$, $\Array$ or $\List$ type.
	This inlining of structures reduces the depth of the schema tree.
\end{enumerate}
After the above preprocessing, each field of the structure in the schema definition corresponds to a node in the schema tree, and each field with $\Array$ or $\List$ type is the parent of all the fields of its element structure in the schema tree.

\begin{figure}[!htb]
    \vspace{-8pt}
	\centering
	\includegraphics[width=0.6\columnwidth]{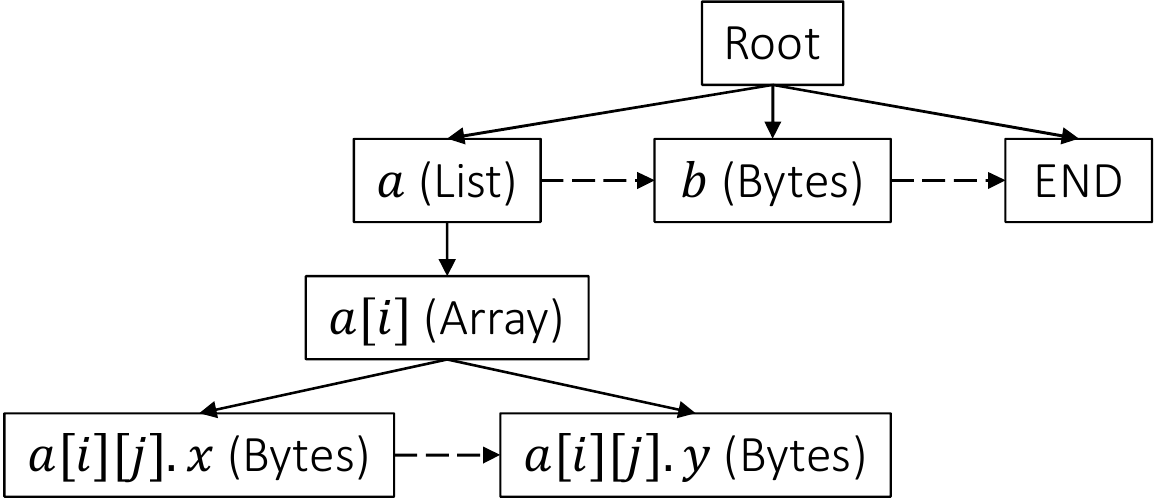}
    \vspace{-7pt}
	\caption{Schema tree for schema in Figure \ref{fig: idl example}} \label{fig: schema tree}
    \vspace{-5pt}
\end{figure}

Figure \ref{fig: schema tree} shows the schema tree of the example schema in Figure \ref{fig: idl example}.
In Figure \ref{fig: schema tree}, solid arrows represent the parent-child relationship, while dashed arrows indicate the ordering between the children nodes with the same parent (i.e., the order of fields in the structure).
We add a special node $\mathsf{END}$ as the last child of the root to signal the end of schema.
All leaf nodes (except $\mathsf{END}$) in the tree are of $\Bytes$ types, while all internal nodes are of either $\Array$ or $\List$ types.
There is no node of a $\Struct$ type because we have inlined all the structures.

\noindent\textbf{Context stack for traversing schema tree:}
The DES logic  generates the token stream by traversing the schema tree in a way similar to pre-order traversal, except that the descendants of an internal node (i.e., fields of the array element or list element) will be visited zero or multiple times depending on the length of the array or list.
A data structure called \emph{context stack} is used to track the necessary information.  
The context stack is a stack, and each entry is called a \emph{context}, which contains information about the array or list that we are deserializing.
All the contexts from bottom to top in the context stack correspond to all the internal nodes on the path from root to the node that we are currently visiting.
For example, when we are visiting node $a[i][j].x$ in Figure \ref{fig: schema tree}, the context stack will contain two contexts.
The bottom context corresponds to node $a$, and the top context corresponds to node $a[i]$.

For a context $C$, we refer to its corresponding internal node as $N$.
The information kepted by $C$ is used to deserialize the array or list represented by $N$, i.e., to traverse the subtree rooted at $N$.
$C$ contains the following fields:
\begin{itemize}
	\item $\Num$: the number of elements that has not been fully deserialized in the array or list of $N$.
	This is the number of traversals that we still need for the subtree rooted at $N$.
	
	\item $\Type$: the type of context $C$,  either $\Array$ or $\List$.
	This is used to output the ending token (i.e., $\mathsf{array\mhyphen{}end}$ or $\mathsf{list\mhyphen{}end}$) after all elements of the array or list of $N$ have been deserialized.
	
	\item $\ChildPtr$: the pointer that points to the first child node of $N$.
	For example, the $\ChildPtr$ of the context for node $a[i]$ in Figure~\ref{fig: schema tree} will point to $a[i][j].x$.
	This pointer is used when we start the next traversal of the subtree rooted at $N$.
	
	\item $\NextPtr$: a pointer to the \emph{next} sibling of $N$, i.e., the node ordered right after $N$ by dashed arrows among all siblings of $N$.
	If the next sibling does not exist, $\NextPtr$ is $\NULL$.
	For example, the $\NextPtr$ of the context for node $a$ in Figure \ref{fig: schema tree}  points to node $b$, while the $\NextPtr$ of the context for node $a[i]$ is $\NULL$.
	$\NextPtr$ indicates which node to visit next in the schema tree when context $C$ ends (i.e., the corresponding array or list has been fully deserialized).
	If the pointer is $\NULL$,  the upper-level context should be consulted.
\end{itemize}

The contexts in the context stack track all the arrays and
lists that are being deserialized, and the context stack is used to control
the traversal of the schema tree.  
The traversal is started by visiting the first child of the root node.  For
convenience, we refer to the node being visited as \VisitNode, refer
to the field in message schema corresponding to \VisitNode{} as
\VisitField{}, and refer to the context at the top of the context
stack as \TopContext.  We take the following actions based on the
types of \VisitNode{} and \VisitField:
\begin{itemize}
	\item If \VisitNode{} is $\mathsf{END}$, then DES is done, and the traversal stops.
	
	\item If \VisitField{} is of type $[\Bytes,n]$, we read the $n$-byte data from the phit stream and emit a token for \VisitField{}.
	If \VisitNode{} is not the last child of its parent, we go to visit its next sibling.
	Otherwise,  deserializing one element of an array or list must be complete, so we decrement the $\Num$ of \TopContext{} by one, and then check whether $\Num$ (after decrement) reaches zero.
	If $\Num$ is greater than zero, we restart deserializing the element of the array or list of \TopContext{} by  visiting the node pointed by the $\ChildPtr$ in \TopContext{}.
	Otherwise, we emit the $\mathsf{array\mhyphen{}end}$ or $\mathsf{list\mhyphen{}end}$ token based on the $\Type$ in \TopContext, check whether the $\NextPtr$ in \TopContext{} is $\NULL$, and pop the context stack.
	If the $\NextPtr$ of the just popped context is not $\NULL$, we visit the node pointed to by this $\NextPtr$.
	Otherwise,  the $\Num$ of the new \TopContext{} is decreased by one, and  the above operations are repeated until  a node to visit is found.
		
	\item If \VisitField{} is of the $\Array$ or $\List$ type, we read the
	number of elements in the array or list from the phit
	stream, and then emits an $\mathsf{array\mhyphen{}length}$
	or $\mathsf{list\mhyphen{}begin}$ token.
	If the number of elements is not zero, we push a new context to the context stack, and visit the first child node of \VisitNode.
	Otherwise, we emit  the $\mathsf{array\mhyphen{}end}$ or $\mathsf{list\mhyphen{}end}$ token.
    The next node to visit is found in the same way as when \VisitField{} is of type $[\Bytes,n]$.
\end{itemize}
The above traversing policy can be implemented using a very small FSM which is \emph{independent} from the message schema.
The size of the context stack is determined by the maximum depth of the schema tree, which should not be very large.

\noindent\textbf{Storing schema tree in hardware:}
The schema tree is stored into hardware by encoding the tree into a ROM, which is referred to as the \emph{schema ROM}.
Each schema ROM entry stores one node in the schema tree, containing information such as the type of the corresponding field, the user-specified tag, etc.
Nodes sharing the same parent will be stored consecutively in the schema ROM and the order is the same as that in the schema tree.
In this way, we only need to increment the schema ROM index by one if we want to visit the next sibling of the current visiting node.
For each node with the $\Array$ or $\List$ type, its corresponding ROM entry additionally contains the index of the schema ROM entry of its first child node.

The combination of the schema ROM, context stack, and traversing FSM can control the consumption of phit stream to emit tokens.
These together constitute the DES logic.

\noindent\textbf{Critical path delay and area cost of the DES logic:}
The critical path delay is insensitive to the
message schema for two reasons.  First, the delays of traversing the
FSM and its context stack are both independent of the message schema.
Second, even though the size of the schema ROM is determined by the
message schema, ROM is a highly optimized IP core and its delay should
also be small.

The area cost is also insensitive to the message schema. 
The area of the traversing FSM is unaffected by the schema.
The size of the context stack is equal to the depth of the schema tree, which should not be large. 
Since ROM is a highly optimized IP core, the schema ROM should not incur large area costs unless the message schema is extremely complex.

\subsection{Implementation of HW-to-SW Messaging}

The implementation of HW-to-SW messaging is very similar to that of
SW-to-HW messaging.  The SER logic in hardware can also be
performed by traversing the schema tree.  The only problem is that the
hardware SER logic cannot behave like the software
SER function, which writes the length
of a list to the phit stream before writing all the list elements.

This problem can be solved by leveraging the fact that software
can buffer the whole message.  That is, the hardware SER logic
writes the length of an array or a list to the phit stream
\emph{after} writing all the elements of the array or list.
Since the whole binary data for the message is buffered in software, the software DES function can simply
deserialize the message from the end of the data buffer.

\subsection{Implementation of HW-to-HW Messaging}

The implementation of HW-to-HW messaging is also similar to the implementations of hardware SER/DES logic in SW-to-HW and HW-to-SW messaging.
However, since neither the sender nor the receiver in HW-to-HW messaging can buffer the whole list, the previous solutions about transferring $\List$ data type are no longer applicable.
Thus, we need a new solution to specify the end of a list under the constraint that both sender and receiver process data in a streaming fashion.

One solution would be adding a new field to the element type of a list to signal the end of the list.
Since we require all data types to be byte-aligned to reduce the logic of manipulating phit and token data, the new field must be at least one-byte large.
This will increase the size of the serialized list at a ratio of $1 / number\ of\ bytes\ in\ element\ type$.
When the user configures a large byte width or the original element type is of small size, the overhead incurred by the new field is significant.

\noindent\textbf{Framing solution for list:}
HGum uses an alternative solution instead of adding new fields.
The idea is to allocate a moderate amount of on-chip buffer in the hardware SER logic to store part of the serialized list data.
When the serialized list data fills up the buffer or the list has been fully serialized, the data inside the buffer forms a \emph{frame}. 
Then the frame is prefixed with a header containing the frame size and is sent to the network.
The hardware DES logic first sees the frame header and then determines how many bytes to consume for the list.

The on-chip buffer can be implemented as a FIFO with an additional
write port to set the frame header.  In this way, after the buffer
is filled up, the old frame can be sent to the network and at the meantime a new frame can start being constructed (i.e., by enqueuing new data into the FIFO).
The buffer size can be configured by the user.  The overhead in the
size of the serialized list is one header per frame, and the overhead
in SER/DES throughput is a few cycles per frame. Thus, the overall
overheads of this framing solution is negligible, as long as the
user configures the buffer size to a not-so-small number (e.g.,
buffer can be 512-phit large since the depth of block RAM is 512 in
Altera FPGA) and the size of the serialized list is large (e.g.,
larger than the buffer size).

\begin{figure}[!htb]
    \vspace{-7pt}
	\centering
	\includegraphics[width=0.8\columnwidth]{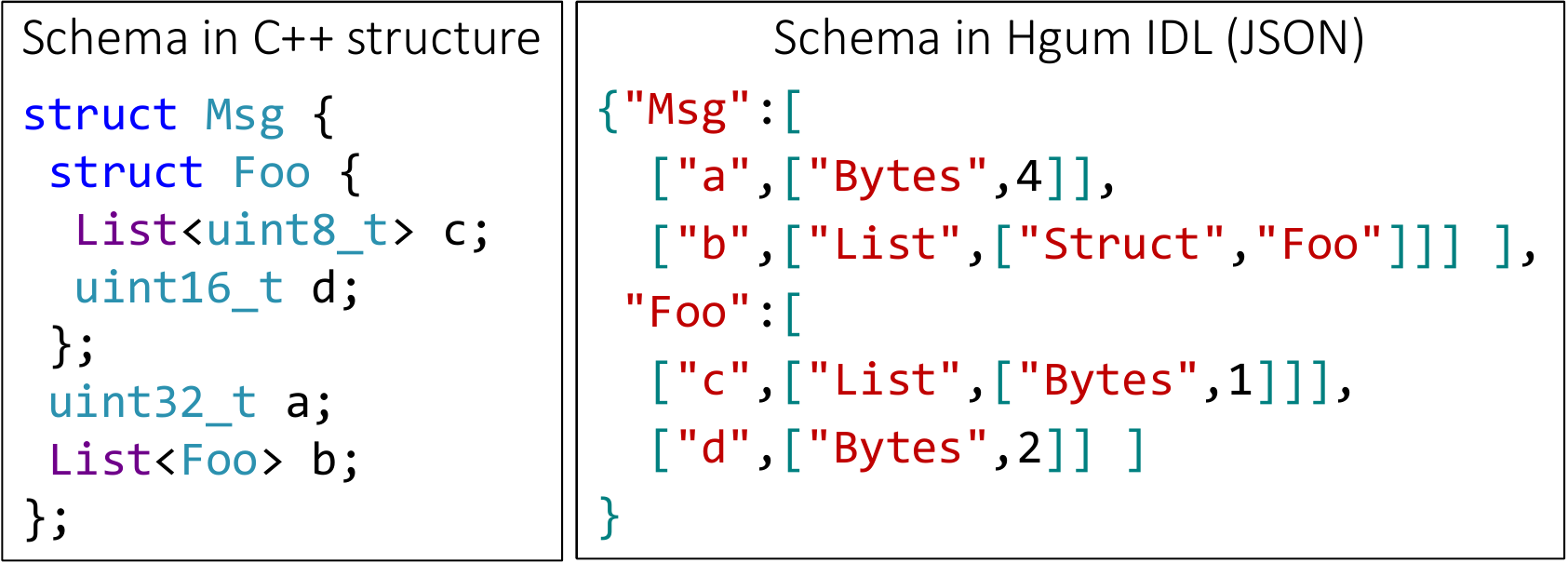}
    \vspace{-7pt}
	\caption{Example schema for framing protocol} \label{fig: frame schema example}
    \vspace{-6pt}
\end{figure}

\noindent\textbf{Details of framing protocol:}
The framing protocol must ensure that the DES logic can unambiguously determine which part of the message schema is encoded in the frame based on the frame header.
This cannot be achieved if the frame header only contains the size of the frame.
To better illustrate the problem, consider an example schema in Figure~\ref{fig: frame schema example}.
The schema is shown in both C++ structure and HGum IDL (JSON).
After the DES logic emits the token for field $a$, there are three possible cases for the rest of the message: 
\begin{enumerate}
	\item List $b$ is an empty list.
	\item List $b$ is not empty but list $c$ in the first element of $b$ is empty.
	\item List $b$ is not empty and list $c$ in the first element of $b$ is not empty.
\end{enumerate}
If the framing protocol does not specify more details about when a
frame is started and ended, the DES logic cannot distinguish between the above three cases, i.e., it cannot figure
out whether it will receive a frame, or what content to be expected in
the next frame.  The framing protocol in HGum eliminates all ambiguity.

In the framing protocol of HGum, an empty frame (i.e., only a frame header) always represents the end of a list, so the SER logic must at least send out one frame for each list.
Furthermore, all the data in a single frame must be under the same level of nested lists, e.g., field $d$ and elements of list $c$ cannot be serialized into the same frame.
Thus, all data in a single frame are under the same $\List$ context.
The frame header contains a new field $\ListLevel$, which is the level of nested lists for the data inside the frame, in addition to the frame size.
For example, the $\ListLevel$ field of the frame that contains the data of $d$ will be one, while the $\ListLevel$ field of the frame that contains the elements of list $c$ will be two.

The DES logic tracks the number of $\List$ contexts in the context stack, i.e., the current level of nested lists during the schema traversal.
There are two conditions for the DES logic to expect to receive a new frame header:
\begin{enumerate}
	\item The context stack does not contain any $\List$ context when the DES logic visits a node with a $\List$ type in the schema tree.
	\item The context stack still contains at least one $\List$ context after the DES logic consumes a whole frame.
\end{enumerate}
When the DES logic receives a new frame header, it first checks whether the number of $\List$ contexts in the context stack is equal to the $\ListLevel$ field in the header.
If it is not, the DES logic will keep traversing the schema tree until equality is reached.
After that, the DES logic must have reached the same node in schema tree as the one where the SER logic starts to generate the frame, so it can construct the right tokens using the frame data.

\section{Evaluation of HGum} \label{sec: eval}

The hardware SER/DES logic generated by HGum runs at high clock
frequency and consumes very little area.  Even for a fairly complex
message schema which contains various combinations of three levels of
nested arrays and lists, the four generated hardware SER/DES modules
for three types of messaging (i.e., SW-to-HW DES, HW-to-SW
SER, and HW-to-HW SER and DES) in the
Altera Stratix V D5 FPGA can all be clocked over 200MHz, and consume
5.5\% of the logic resources and less than 0.2\% of the block RAMs in
total.

Besides the frequency and area, we next evaluate the throughput of the generated logic and the usability of HGum.

\subsection{Throughput of Generated SER/DES Logic} \label{sec: eval array list}

\begin{figure}[!htb]
    \vspace{-5pt}
	\centering
	\includegraphics[width=0.75\columnwidth]{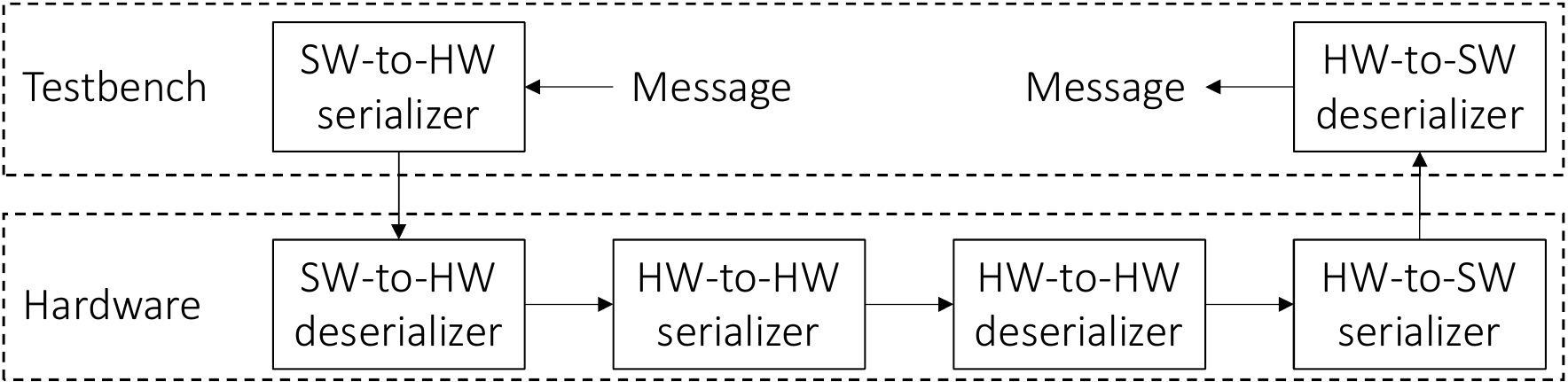}
    \vspace{-5pt}
	\caption{Simulation microarchitecture for evaluating throughput} \label{fig: sim arch throughput}
    \vspace{-5pt}
\end{figure}

We evaluate the throughput of the hardware SER/DES logic generated by HGum in simulation using the microarchitecture shown in Figure~\ref{fig: sim arch throughput}.
This microarchitecture implements the loopback of a message: the message is transferred first from software to hardware, then to another hardware, and finally back to software.
The SER/DES modules in hardware are automatically generated by HGum, while the SER/DES functions in software are replaced by testbench modules, which are SystemVerilog classes that perform SER and DES.
These SystemVerilog classes are also generated by HGum.

We measure the throughput of this loopback system in terms of the number of messages processed per cycle.
The phit in the system is set to 128-bit (16-byte) wide, and the maximum size of a frame in the HW-to-HW SER logic is set to 500-phit large.
We use the following two simple message schema in this evaluation:
\begin{enumerate}
	\item An array of 128-bit data. 
	That is, the schema only contains one field with type $[\Array,[\Bytes,16]]$.
	The client schema specifies that no $\mathsf{array\mhyphen{}end}$ token should be outputted by any DES logic.
	\item A list of 128-bit data. 
	That is, the schema only contains one field with type $[\List,[\Bytes,16]]$.
\end{enumerate} 
In the ideal case, each hardware SER/DES module should consume or emit
one token per cycle. We use this criterion to calculate the optimal
throughputs for the above message schema, and compare the measured
throughputs of HGum against the optimal values.

\begin{figure}[!htb]
    \vspace{-5pt}
	\centering
    \subfloat[Array\label{fig: array throughput}]{\includegraphics[width=0.48\columnwidth]{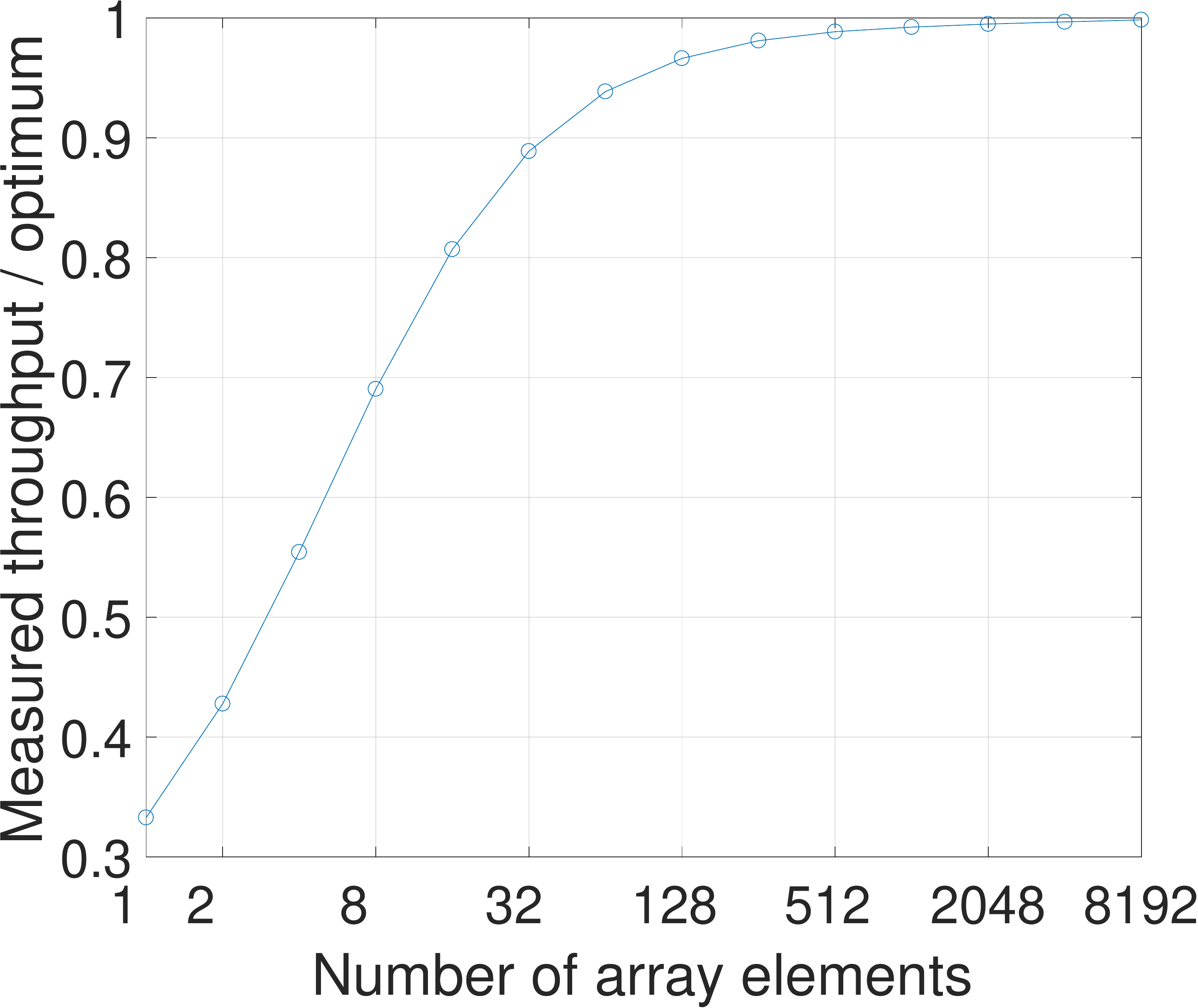}} \quad   
	\subfloat[List\label{fig: list throughput}]{\includegraphics[width=0.48\columnwidth]{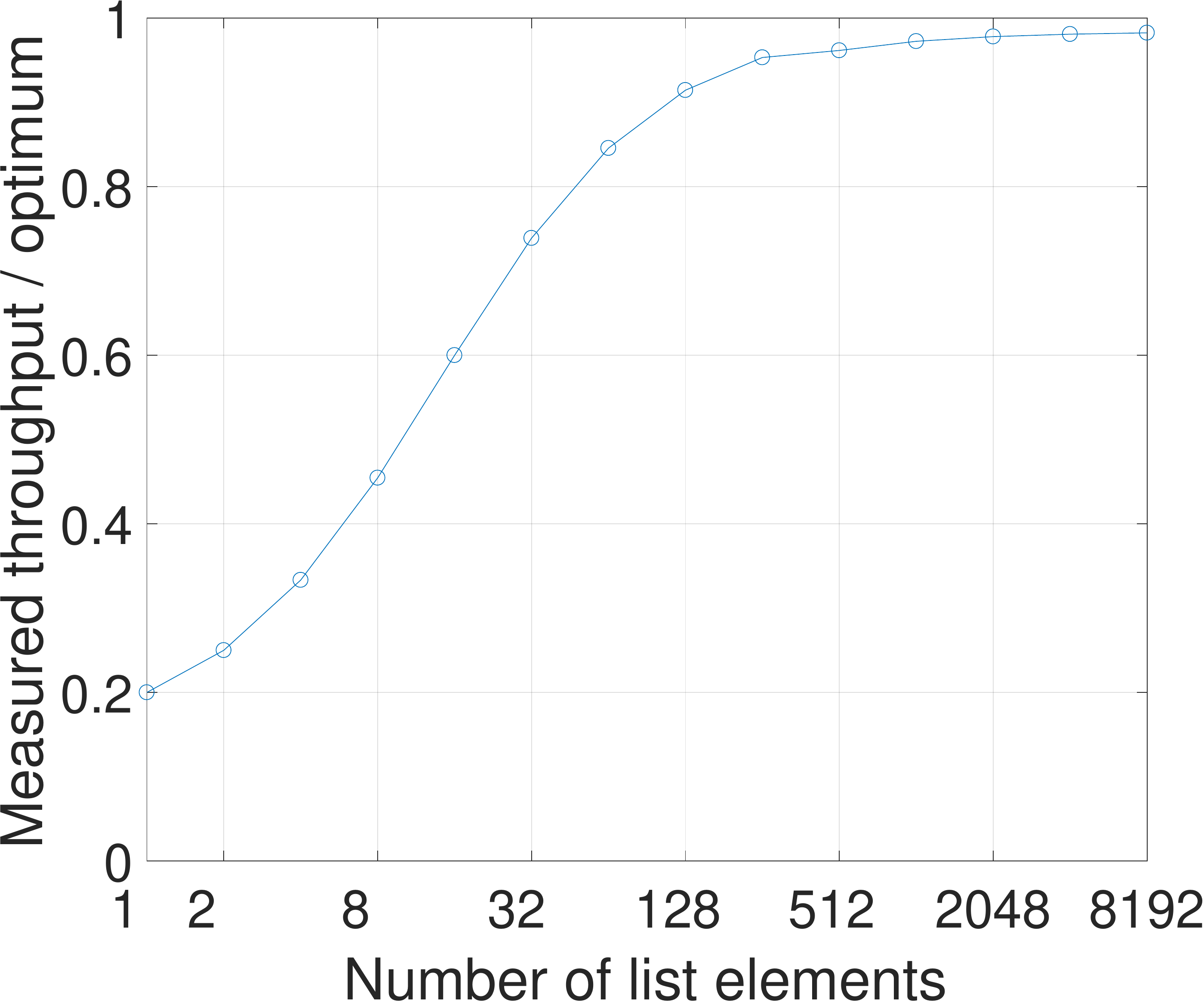}}    
    \vspace{-2pt}
    \caption{Relative throughput for array and list of 128-bit data against optimum}
    \vspace{-8pt}
\end{figure}

\noindent\textbf{Throughput for array of 128-bit data:}
Since an $n$-element array corresponds to $n+1$ tokens, the optimal throughput is $1/(n+1)$ messages per cycle.
Figure~\ref{fig: array throughput} shows the ratios of the measured throughputs over the optimal throughputs for a message schema which is an array of 128-bit data, when the array length changes from 1 to 8192.
The overhead of generated hardware SER/DES logic is large for small array lengths because the SER/DES logic requires a few extra cycles to process the array.
This overhead will diminish to almost zero as the array length increases.

\noindent\textbf{Throughput for list of 128-bit data:}
Since an $n$-element list corresponds to $n+2$ tokens, the optimal throughput is $1/(n+2)$ messages per cycle.
Figure \ref{fig: list throughput} shows the ratios of the measured throughputs over the optimal throughputs for a message schema which is a list of 128-bit data, when the list length changes from 1 to 8192.
Similar to the case of array schema, the overhead is large for small list lengths and will decrease as the length increases.
However,  throughput cannot reach the optimal value even for very long lists, because the SER/DES logic has a overhead of a constant number of cycles for each frame.
The frame in the evaluation can at most hold 500 phits, which is already large enough to make the overhead for long lists negligible.

\noindent\textbf{Summary:}
The SER/DES logic generated by HGum can achieve near-optimal throughputs for long arrays and lists, but it may incur substantial overheads for very short arrays and lists.

\subsection{Real Case Study of HGum}

As a real use case of HGum, we ported the Feature Extraction (FE)
accelerator, which is the first stage of the 8-stage FPGA pipeline
that accelerates the Bing Ranking algorithm~\cite{putnam2014reconfigurable}, 
to the HGum framework.
For convenience, we refer to the original FE accelerator
as \textbf{FE-orig}, while we refer to the new FE accelerator using
HGum as \textbf{FE-HGum}.

\begin{figure}[!htb]
    \vspace{-7pt}
    \centering
    \begin{minipage}[b]{0.48\columnwidth}
        \centering
        \includegraphics[width=\columnwidth]{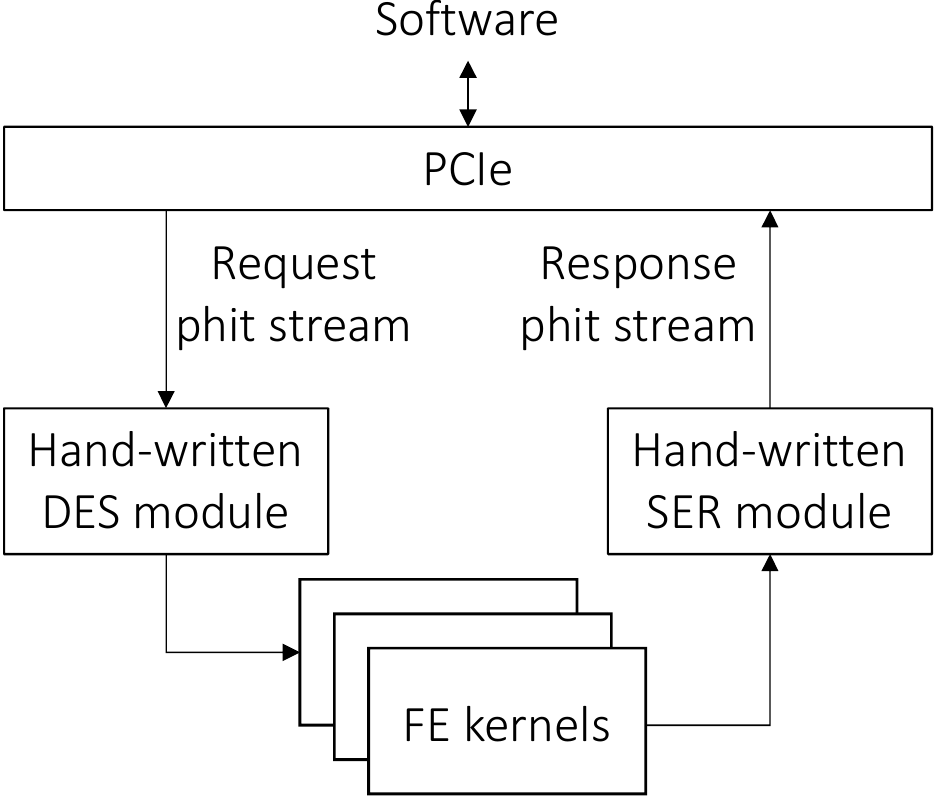}\vspace{-8pt}
        \caption{FE-orig microarchitecture}\label{fig: orig fe}
    \end{minipage}\quad
    \begin{minipage}[b]{0.48\columnwidth}
        \centering
        \includegraphics[width=\columnwidth]{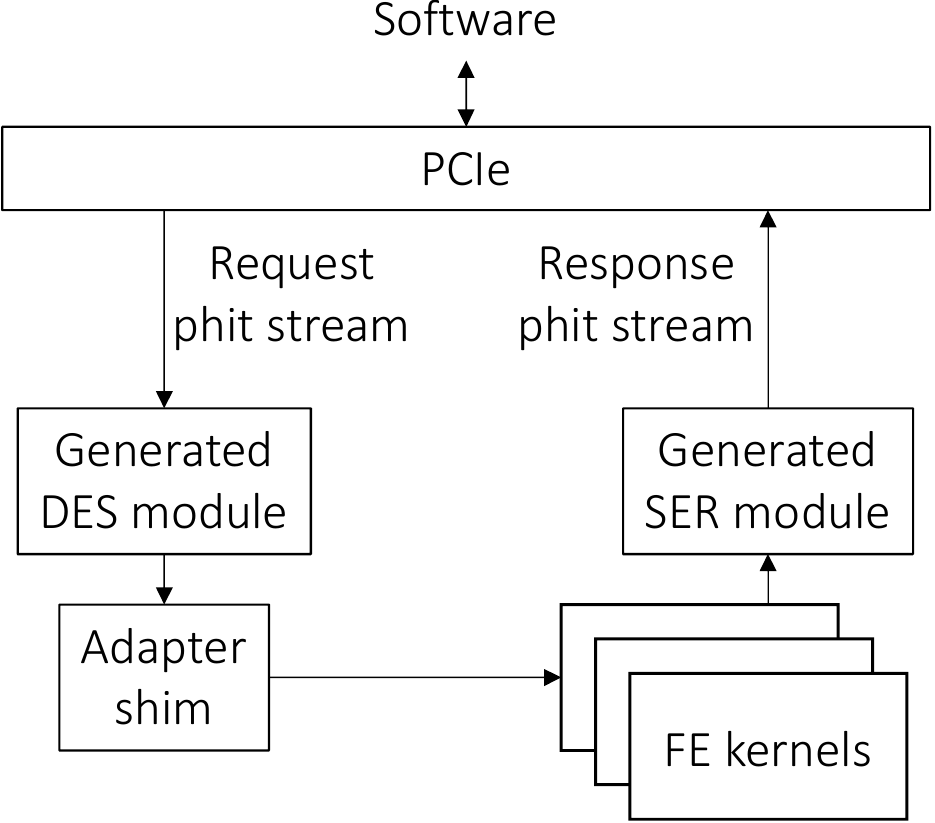}\vspace{-8pt}
        \caption{FE-HGum microarchitecture}\label{fig: hgum fe}
    \end{minipage}
    \vspace{-15pt}
\end{figure}

The FE accelerator is implemented on a single FPGA, and it communicates with the host software via PCIe.
Figure~\ref{fig: orig fe} shows the hardware microarchitecture of FE-orig, which consists of hand-written SER/DES logic and computation kernels to extract features.
The schema of the request message from software to hardware contains multiple levels of nested arrays and structures.
That is, the element type of an array in the schema is a structure that contains other arrays as structure fields.
Due to the complexity of the schema, the hand-written DES logic is implemented using a complicated FSM, which requires huge engineering and verification efforts.
The response message contains a list of extracted feature and another list of meta information, which is relatively easier to serialize.

We ported FE-orig to the HGum framework (i.e., implemented FE-HGum) in one man-week.
The microarchitecture of FE-HGum is shown in Figure~\ref{fig: hgum fe}, where SER/DES logic is automatically generated by HGum.
The hand-written SER logic in FE-orig is now completely replaced by the automatically generated one.
As for deserialization, we need to manually construct an adapter shim module to connect the generated DES logic to the FE kernels, because the input format of the FE kernels is not the same as the token stream outputted by the generated DES logic.
In the user-defined tag for each token (i.e., the client schema), we encode information on how to convert such a token to an input to the FE kernel.
As a result, the adapter shim can be implemented using very simple logic.
The number of lines of codes of the adapter shim module (the only hand-written logic for DES in FE-HGum) is merely 27\% of that of the hand-written DES logic in FE-orig.

FE-HGum can be synthesized at the same clock frequency as FE-orig.
In terms of area, FE-HGum costs relatively 3.4\% more logic while 12\% less block RAMs than FE-orig does.

\begin{figure}[!htb]
    \vspace{-5pt}
	\centering
	\includegraphics[width=0.65\columnwidth]{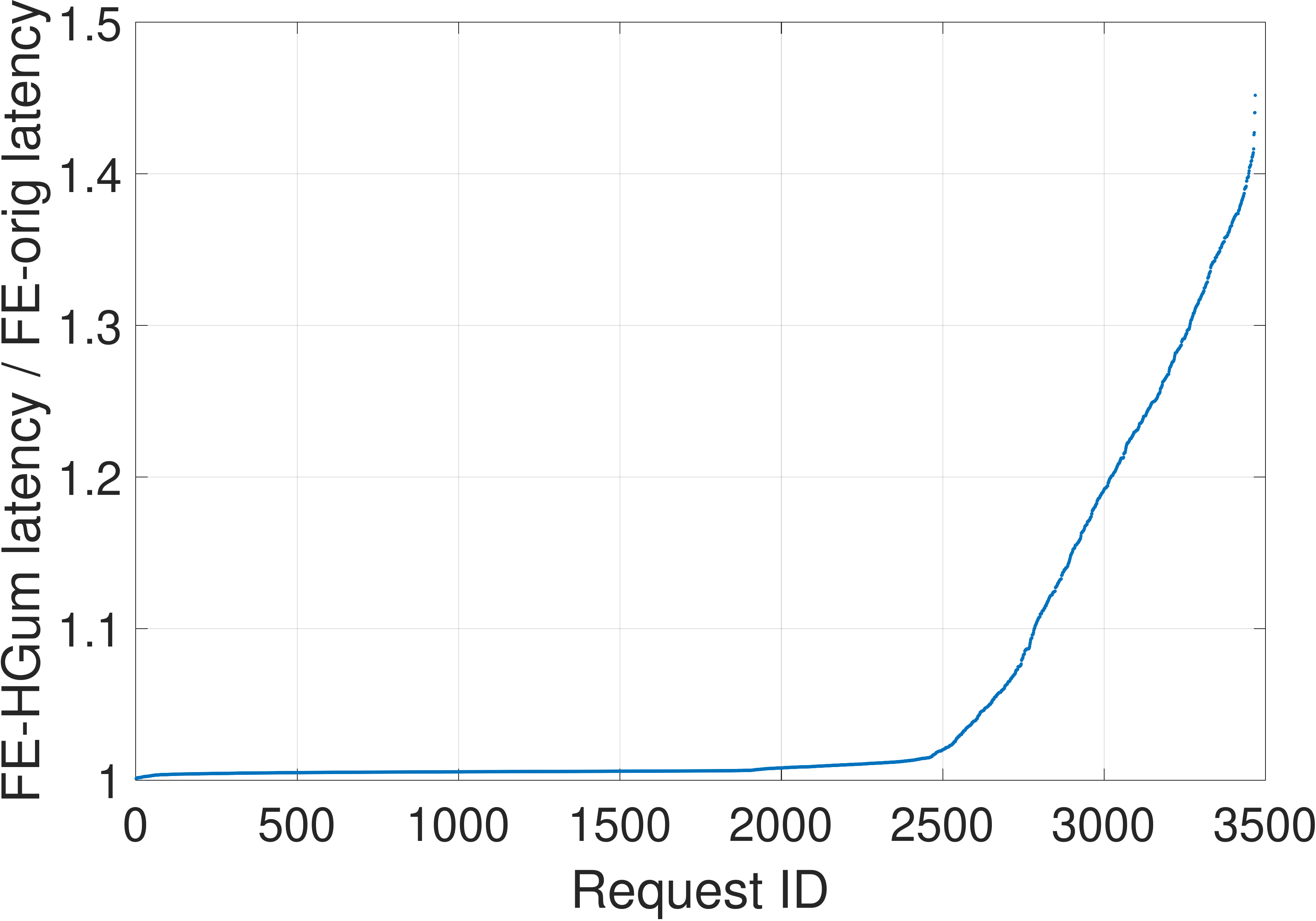}
    \vspace{-5pt}
	\caption{Relative latency for FE-HGum against FE-orig} \label{fig: fe lat}
    \vspace{-10pt}
\end{figure}

As for performance, since FE is a blocking operation (i.e., only one
request can be processed on FPGA), we measured the latency from the
start of hardware deserialization to the end of hardware serialization
for each request.  We have tested 3468 requests on FPGA. All
requests are intercepted from real Bing Ranking service traffic. 
The ratio of the latency of
FE-HGum over that of FE-orig for each request is shown in Figure~\ref{fig: fe lat}.
The latency of FE-HGum is generally higher than
that of FE-orig, because the SER/DES logic generated by HGum requires
additional cycles to process each array as we have analyzed in Section~\ref{sec: eval array list}.
Despite these extra cycles, the latency
of FE-HGum is almost the same as that of FE-orig for most requests.
The geometric mean of the latency ratios is 1.05, i.e., FE-HGum incurs
merely 5\% latency overheads on average.

In summary, porting FE to HGum illustrates that HGum is easy to use
and can save significant engineering efforts by reducing a large amount
of hand-written codes for deserialization.  Furthermore, the quality
of the hardware generated by HGum is almost the same as that of the
hand-written implementation.

\section{Conclusion} \label{sec: conclude}

In the paper we have presented HGum, the first messaging framework for hardware (FPGA) accelerators that handles the serialization and deserialization of large and complex messages for all three directions of communication among hardware and software (i.e., SW-to-HW, HW-to-SW and HW-to-HW). 
By leveraging novel techniques in streaming processing, algorithmic schema parsing and flexible message tokenizing, HGum can generate high-performance and low-cost SER/DES logic. 
Our evaluation of HGum demonstrates that HGum can not only reduce significant amounts of engineering efforts, but also generate hardware with almost the same quality as ad-hoc manual implementations.

\section{Acknowledgement}
We thank all the reviewers for their helpful feedback, and the Microsoft Catapult team for their help on this project.
Sizhuo Zhang was an intern at Microsoft during this project.

\bibliographystyle{IEEEtran}
\bibliography{ref}  

\end{document}